# Morphological Evolution of Nickel–Fullerene Thin Film Mixtures

Giovanni Ceccio[1,*], Kazumasa Takahashi[2], Romana Miksova[1], Yuto Kondo[2], Eva Stepanovska[1], Josef Novak[1], Sebastiano Vasi[3,4,*], Jiri Vacik[1]

1. Department of Neutron and Ion Methods, Nuclear Physics Institute of Czech Academy of Science, Hlavni cp. 130 Husinec - Rez, 25068, Czech Republic
2. Department of Electrical Engineering, Nagaoka University of Technology, 1603-1 Kamitomioka, Nagaoka, Niigata, 940-2188, Japan
3. Department of Mathematics and Computer Sciences, Physical Sciences and Earth Sciences, University of Messina Viale F. Stagno d'Alcontres 31, 98166, Messina, Italy
4. OPENFIS S.R.L.-SPIN OFF ACCADEMICO, University of Messina, LABORATORIO A2AT3, Viale F. Stagno D'Alcontres 31, 98166 Messina, Italy

* Correspondence: ceccio@ujf.cs.cz; vasis@unime.it

**Abstract**

Hybrid systems consisting of metal–fullerene composites exhibit intriguing properties but often suffer from thermal instability. With proper control, such instability can be harnessed to enable the formation of sophisticated nanostructures with nanometric precision. These self-organization phenomena are not limited to thermal stimulation alone but can also be triggered by other external stimuli. In this work, we investigate the morphological evolution of thin films composed of evaporated C60 and sputtered nickel mixtures, focusing on how external stimuli influence both their structural and electrical properties. Thin films were prepared under controlled deposition conditions, and their surface morphology was analyzed using advanced characterization techniques. Progressive changes in film morphology were observed as a function of composition and external treatment, highlighting the interplay between metallic and molecular components. In particular, it was observed that, due to the annealing treatment, the sample undergoes strong phase separation, with the formation of structures tens of microns in diameter and an increase in electrical resistance, exhibiting insulating behavior. These findings provide insights into the mechanisms governing hybrid thin film formation and suggest potential applications in electronic, optoelectronic, and energy-related devices.

**Keywords:** Nanostructures, fullerenes, thin films, ion beams

## 1. Introduction

Nanostructured materials can be synthesized using a wide variety of methods and approaches, enabling fabrication in a broad range of forms and combinations. [1-3]. For these reasons, nanomaterials are continuously produced and investigated due to their versatility and remarkable properties, both for fundamental sciences and advanced technological applications [4-7]. An interesting class of nanomaterials with great potential is based on mixture of metal- and carbon-based nanostructures [8, 9]. The prospects of using such nanostructures lie, for example, in electron-spin-related applications and in the electronic flexibility of the metal–organic interface, whose proper understanding opens broad opportunities for future technologies [10, 11]. The possible structures that can be formed range from zero- to three-dimensional (0D to 3D), with diverse morphologies and sizes. Within these broad possibilities range, the choice of $C_{60}$ fullerene has attracted considerable attention in these hybrid systems due to the remarkable features of this allotropic form, including mechanical stability, electrical conductivity and optical characteristics [12,13]. Furthermore, coupling $C_{60}$ with transition metals further enhances the functionality of the system, offering opportunities to tune its electronic properties, especially when magnetic elements such as Ni or Co are employed [14–16]. For example, such an assembled Ni–$C_{60}$ system holds great potential in spintronics, owing to the remarkable organization of magnetic Ni clusters within the deposited mixture and the robustness of $C_{60}$ [17]. Unfortunately, these hybrid systems are generally considered thermodynamically and structurally unstable, mainly due to the high internal stress resulting from the mixing of largely immiscible phases and the vulnerability of $C_{60}$ molecular cages to photo-oxidation, polymerization, or fragmentation in environments with strong catalytic activity [17,18]. Nevertheless, such instability can be exploited for our purposes by implementing well-defined modification protocols that enable controlled formation of self-organized structures and deliberate tuning of electrical properties. The post-deposition treatments examined in this study—vacuum annealing, pulsed laser irradiation, continuous Ar ion irradiation, and pulsed C ion irradiation—were selected to probe complementary physical mechanisms influencing the evolution of composite films. Vacuum annealing primarily activates thermally driven diffusion and phase separation under near-equilibrium conditions, serving as a reference for thermodynamic instability. Pulsed laser irradiation introduces transient and localized thermal excitation, granting access to non-equilibrium pathways and stress-induced self-organization phenomena. Continuous Ar ion irradiation emphasizes energy transfer combined with defect generation and atomic mixing, whereas pulsed C ion irradiation induces similar modifications but with temporal modulation of energy delivery, thereby reducing thermal effects while simultaneously enhancing the likelihood of surface cluster formation. This work aims to elucidate the effects of these post-deposition treatments on films fabricated under identical deposition conditions. All samples were produced simultaneously within the same deposition run using identical source materials and process parameters; thus, any observed differences among the samples can be attributed exclusively to the applied external treatments rather than to intentional compositional variations. The prepared samples were subjected to vacuum annealing, pulsed laser irradiation in air, and ion irradiation (by both continuous and pulsed beams). The analyses revealed that the hybrid systems evolved in distinct ways, exhibiting variations in their nanostructures and macrostructures depending on the external perturbation applied. Some previous studies investigate effect of only one post-production treatment, often on films fabricated with different characteristics. As a result, it remains not fully understood how the external excitation itself, rather than differences in composition, controls the evolution of metal fullerene composite films. The novelty of the present study lies in the systematic comparison of distinct post-deposition treatments on the same nominal composite films.

## 2. Materials and Methods

The investigated systems have been produced by means of simultaneous deposition of Ni and $C_{60}$ onto Si crystal substrate (100) at room temperature. Prior to the film's growth, the substrates' surfaces were cleaned with acetone and isopropyl alcohol. After this, a drying step using nitrogen gas was performed. The deposition was performed in laboratory made multipurpose Low Energy Ion Facility (LEIF), dedicated to the fabrication and modification of nanomaterials, especially in thin film form [19]. For the preparation of the hybrid matrix, a high purity Ni target (Kurt J. Lesker, 2.0" Dia. x 0.125" Thick 99.9 % purity) was sputtered by means of an intense Ar ion beam produced by a duoplasmatron ion source and focused on a multipurpose vacuum chamber. At the same time, $C_{60}$ fullerene (99.9% nanografi) was evaporated from a quartz crucible present in the vacuum chamber inside pure W coil. The deposition of Ni was performed at 1 mA of ion current accelerated at 20 keV and with a beam spot of 3 cm$^2$ on target, the angle between beam and normal to the target was kept 60° by a rotating holder. The crucible, filled with fullerene, was kept at the temperature of 450°C by means of a digital temperature controller and a thermocouple type K placed inside the crucible. During the beam calibration and heat up procedure of the crucible, a shutter is employed to avoid unwanted depositions on the substrates. The deposition was performed with a base pressure of 4.3E-6 mbar for a duration of an hour. The elemental composition of the deposited films was analyzed by Rutherford Backscattering Spectroscopy (RBS) available at Tandetron Laboratory of NPI. The measurements were carried out using a 2.0 MeV He$^+$ ion beam produced by the 3MV Tandetron accelerator working within CANAM infrastructure [20]. RBS identify the heavy elements within a sample by analyzing the residual energy of ions backscattered from specific depths [21]. The backscattered He$^+$ ions were recorded by an Ultra-Ortec PIPS detector positioned at a scattering angle of 170°, providing an energy resolution of approximately 20 keV. During the measurements, the ion beam current was maintained at about 5 nA and the beam spot size was roughly 1 mm$^2$. Experimental spectra were analyzed using the SIMNRA software package to determine the elemental depth profiles [22]. To analyze the possibility of tuning the produced nanostructures, several external stimuli were applied. LEIF facility was used to irradiate the selected samples with continuous Ar beam. For the irradiation, the Ar beam intensity was reduced to 100 µA/cm$^2$ and accelerated at the same energy of 20 keV. The produced beam was irradiating the samples perpendicularly to the surface, with a total dose of 1E15 Ar/cm$^2$. During the irradiation, the samples temperature did not increase due to the short time of irradiation (approx. 16 seconds per sample). Additional samples were irradiated with pulsed C beam at the same energy and fluence using a Laser Ion Source lab-made at Nagaoka University of Technology (NUT). The used laser source was a Quantel Brillant operating at 10Hz repetition rate, 6ns pulse duration, 532 nm wavelength and laser energy of 130 mJ. The laser pulse was focused (in vacuum, 7E-5 mbar) on the surface of a C target held at a voltage of 20kV and facing the grounded samples. To reach the same fluence used for the Ar beam, a total of 12000 pulses were needed for each sample. The needed pulses were calculated after beam calibration using FC placed in samples position. The samples during irradiation were monitored by means of a fast oscilloscope connected to the holder, in order to observe the stability of the irradiating pulsed beam. The same laser light was used to irradiate another set of samples. Lower laser fluence (approx. 5 mJ/cm$^2$) was used to illuminate the surface to induce modification. In particular, the irradiation took place in air, with a total laser pulses of 1000 shots per sample. Finally, the last set was annealed in high vacuum condition for 5 h at 300°C. After these procedures, pristine and modified samples were investigated for their morphological evolution using a Hitachi SU8230 Scanning Electron Microscope (SEM).

Raman spectroscopy analyses were carried out using a LabRAM HR-EVO Horiba spectrometer equipped with a 532 nm laser, a 50× LWD objective, and a CCD Syncerity Horiba detector. The laser power was maintained at 0.8 mW/µm$^2$ to avoid surface destruction of the fullerene. Changes in electrical resistance between pristine and treated samples were examined using a Keithley 2182A Nanovoltmeter in combination with a 6221 DC current source, in a galvanostatic configuration. The measurements were performed on the sample surface under ambient conditions, employing the standard two-point probe method.

## 3. Results

The elemental composition of the films produced in LEIF was investigated after deposition using Rutherford Backscattering Spectroscopy (RBS) available at Tandetron Laboratory. The compositional analysis relies on stopping power and energy loss from backscattered particles to evaluate the depth distribution. Heavy elements such as Ni are particularly well suited to this technique, and possible contamination can be readily detected. Experimental results obtained using 2 MeV He are shown in Figure 1 with the spectra simulated using the SIMNRA software. Representative elemental depth profiles obtained from SIMNRA fits are shown in Figure 1, confirming a homogeneous distribution of Ni throughout the film thickness. These spectra indicate that, for both samples—with and without fullerene—the Ni films are uniform in depth and free of impurities. The calculated thickness from experimental results is 44 nm for the pure Ni film, while for the composite films the total thickness is approximately 38 nm. RBS analysis of the composite film reveals a reduced Ni content of about 40 at.% together with approximately 35 at.% oxygen, with the remaining fraction attributed to carbon. Based on these values, an effective Ni:C atomic ratio of approximately 2:1 is estimated. Due to the intrinsically low backscattering cross section of light elements, direct quantification of carbon by RBS was not feasible at the employed energy.

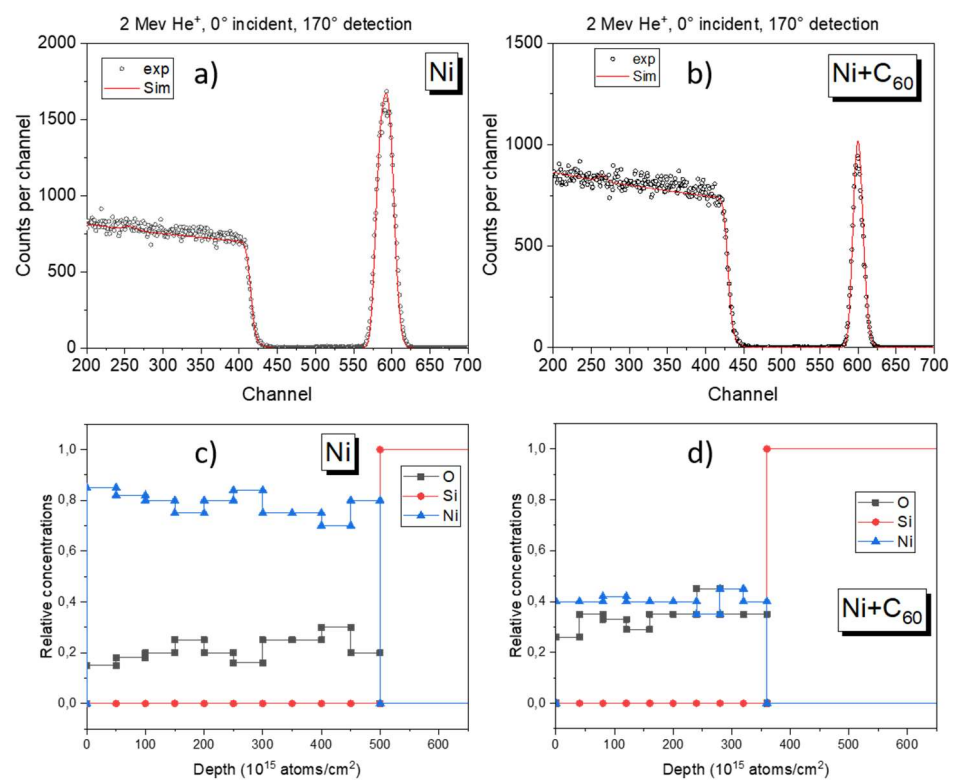

*Figure 1* Panels a) and b) show RBS results on pure Ni and Ni - $C_{60}$ samples obtained with 2MeV He. Black dots are experimental data while red lines are simulation results. Bottom panels c) and d) show the relative concentration of elemental depth profiling for which it is to observe a uniform depth distribution of Ni in both cases.

In this work, the carbon contribution in the composite films was therefore estimated indirectly by combining the measured densities obtained from RBS measurements with an independent calibration of the $C_{60}$ deposition rate, performed under identical evaporation conditions prior to co-deposition. This approach provides a reasonable estimate of the effective Ni:C ratio, while acknowledging the limitations of RBS sensitivity to light elements.

SEM analyses were performed to observe the evolution of the morphology of the films through the modifications performed. Selected SEM micrographs are shown in Figure 2. In the SEM micrograph of pristine sample, a smooth surface and uniform particles distribution is recognized, showing a uniform distribution and deposition. Modified samples exhibited different responses and surface morphologies. The annealed sample shows a large microstructure formed after annealing, where several carbon nucleation creates a regular shape. This represents the release of stress due to the prolonged supply of energy to the system. The stress release occurred symmetrically from the point of origin, resulting in a symmetric structure. In addition, the dimensions suggest an agglomeration of several smaller points merging into a larger one.

The sample that was irradiated with the laser shows the same nucleation points, with sizes around 1μm, but they are much smaller and aligned. The energy supplied in a pulsed manner likely allowed stress release, but also in a discontinuous way over time, preventing the nucleation points from merging into a larger one. Of particular interest is the formation of a self-organized pattern, representing the development of a minimum-energy structure. The SEM obtained from the samples irradiated with ion beams -- pulsed or continuous – do not indicate the formation of nucleation points. In this case, probably, these treatments just support the mixing of the materials by developing an amorphous compound on the surface of the films.

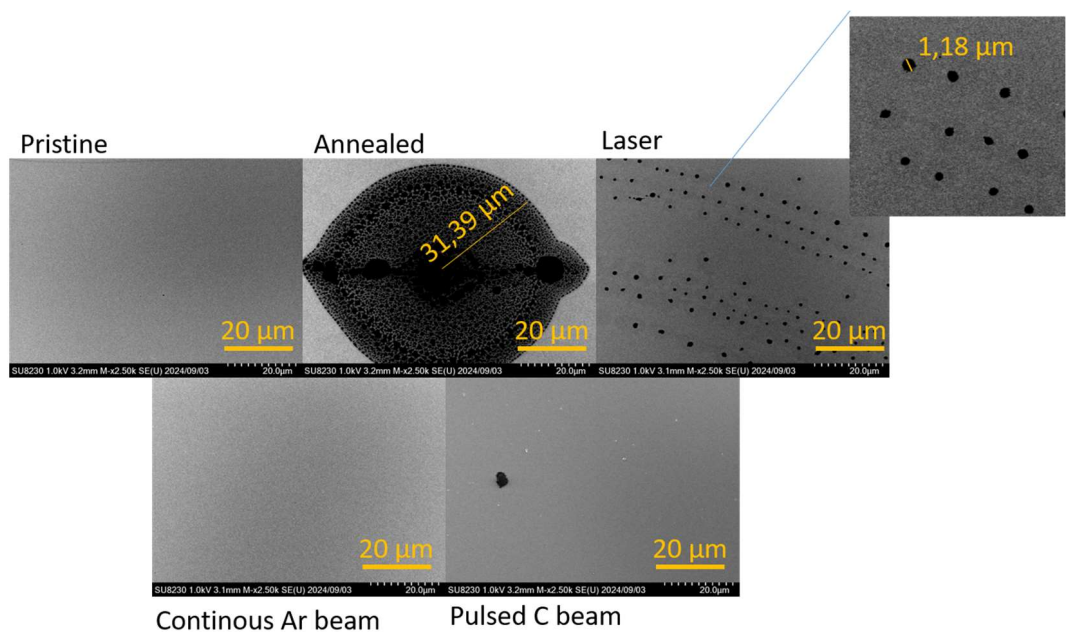

*Figure 2* SEM micrograph of as prepared samples and modified.

Figure 3 illustrates the Raman spectra of pure and metal-fullerene composites materials deposited on a silicon substrate, compared with the vibrational fingerprints of the $C_{60}$ powder used as the precursor for film fabrication, recorded under 532 nm laser excitation. The $C_{60}$ powder exhibits vibrational fingerprints (black trace) in close agreement with values reported in the literature [23, 24]. Together with the Raman modes of Hg symmetry (1100 cm$^{-1}$ (Hg(5)), 1250 cm$^{-1}$ (Hg(6)), 1424 cm$^{-1}$ (Hg(7)), and 1581 cm$^{-1}$ (Hg(8))), the characteristic peak at 1462 cm$^{-1}$ is found. It corresponds to the Ag(2) mode, commonly referred to as the pentagonal pinch, which results from the contraction of pentagonal rings and expansion of hexagonal rings in the carbon framework. These distinct peaks are superimposed on broader bands located between 1330–1360 cm$^{-1}$ and 1560–1600 cm$^{-1}$, known as the D and G bands, respectively—features typical of carbon-based materials under visible-light excitation [25]. The Raman spectrum of the $C_{60}$ film on silicon (Si) substrate (grey trace in Figure 3), shows a sharp tail diminishing at 1065 cm$^{-1}$ which corresponds to the second-order Raman peak of the Si, that, for instance, is present in all the spectra of the deposited hybrid materials. Furthermore, the feature at 1270 cm$^{-1}$ is attributed to a combination of antisymmetric C–C inter-ring stretching and antisymmetric bending of C–H groups formed on the fullerene surface [26]. The 1332 cm$^{-1}$ peak, characteristic of diamond-like carbon materials, is also revealed. A prominent peak at 1450 cm$^{-1}$ is observed, indicative of a photo-transformed state of $C_{60}$ and the formation of polymer chains, which readily occurs during laser irradiation in Raman excitation [23], while the Ag(2) peak at 1462 cm$^{-1}$, characteristic of the $C_{60}$ powder, nearly disappears as it is overshadowed by the tail of the intense 1450 cm$^{-1}$ band. Furthermore, the peak at 1528 cm$^{-1}$, is revealed and it is due to the Gg mode that is visible overall when the fullerene local symmetry breaks [27]. When the $C_{60}$ is co-deposited with nickel across different types of processes, the Raman spectroscopy of the samples subjected to different post-treatments reveals distinct structural modifications induced by these processes. The pristine sample - reported as olive colored line in Figure 3 - exhibits well-defined D (1353 cm$^{-1}$) and G (1590 cm$^{-1}$) bands, characteristic of disordered or partially ordered sp$^2$ carbon domains, likely coexisting with fullerene-like structures, although no distinct fullerene bands are recognizable.

Upon laser irradiation (red line), both bands become slightly narrower and more intense, indicating partial crystallization of amorphous carbon into graphite, likely promoted by localized heating and the catalytic action of nickel for graphite/graphene formation. In contrast, the annealed sample (dark yellow line) displays a strongly reduced Raman signal with broad, featureless features, suggesting thermal degradation or coalescence of $C_{60}$ molecules into amorphous carbon. The samples treated with carbon and argon ion beams (blue and magenta lines in Fig. 3) predominantly display amorphous carbon bands, yet they still preserve weaker but clearly discernible fullerene modes at 1450 cm$^{-1}$, indicating the persistence of distorted $C_{60}$ cages. In contrast, the Ni–$C_{60}$ sample subjected to annealing (dark yellow trace) is the only one that exhibits unequivocal evidence of fullerene Raman modes, with the 1450 cm$^{-1}$ peak being particularly prominent.

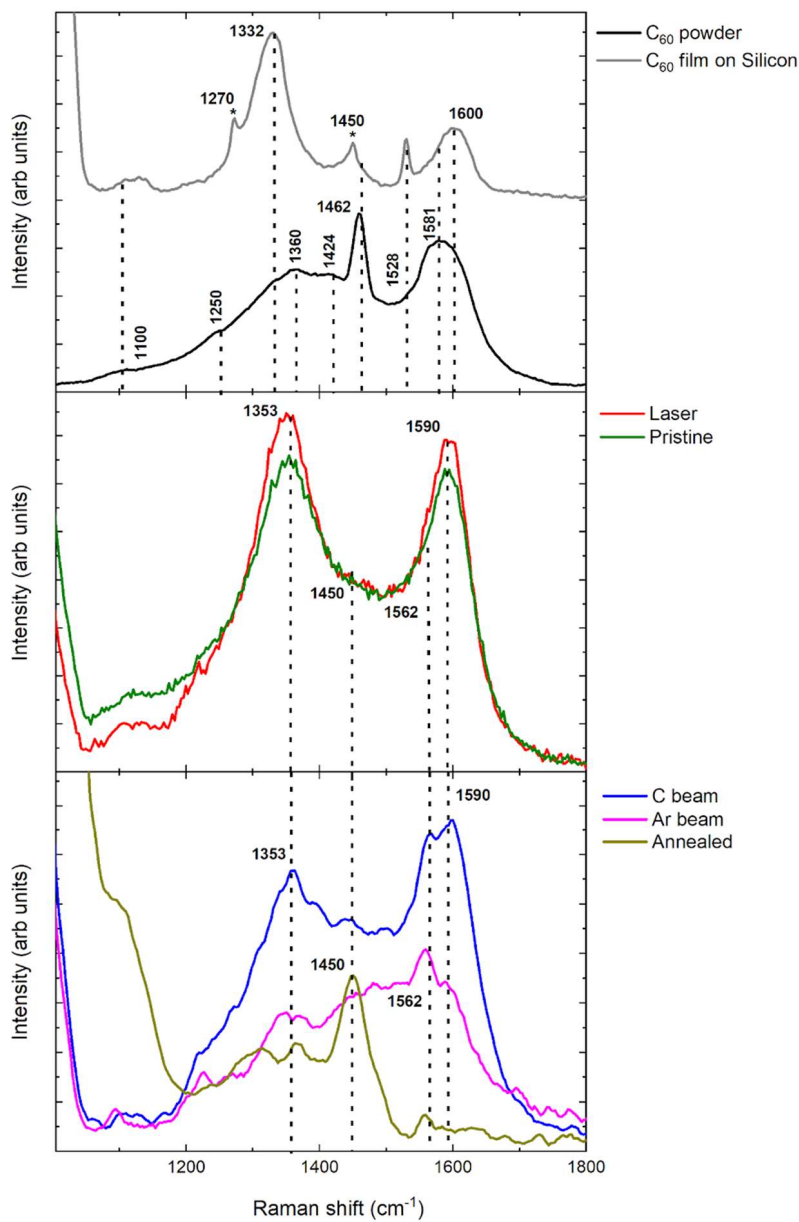

*Figure 3* Raman spectra of $C_{60}$ powder (black line), $C_{60}$ film deposited onto silicon substrate (grey line), and hybrid Ni-$C_{60}$ films, as deposited (olive line) onto silicon substrate and post-treatments: pulsed laser annealing (red line), Carbon (blue line) and Argon (magenta line) beam irradiations and thermal annealing (dark yellow line).

The results concerning the electrical resistance measurements performed on the films are shown in Fig. 4. The annealed sample exhibits a marked increase in effective resistance, indicating strong suppression of electrical transport. It is shown that the resistance of the pristine films is lower, probably due to a good homogeneity of the metallic phase. According to the Raman Analysis, the sample irradiated with laser formed graph-ite structure, and this is also compatible with the low resistance measured. With the amorphization of the film, the resistance increases, for the sample annealed that shows definitely an insulator behavior according also with the Raman results.

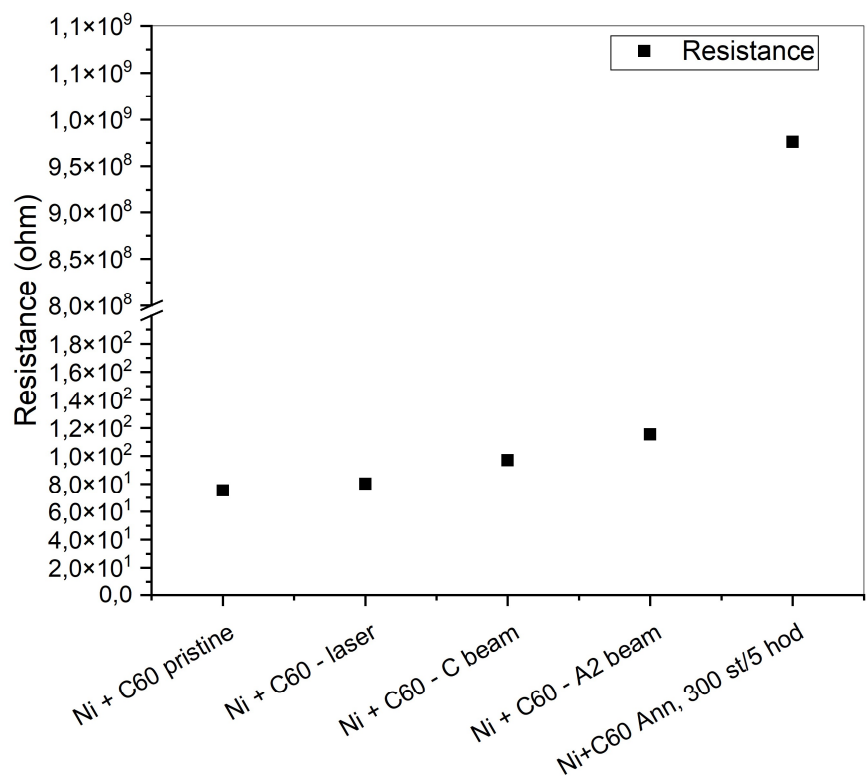

*Figure 4* Results of I-V measurement and calculated electrical resistance.

## 4. Discussion and conclusions

The systematic study presented in this work exploits a highly versatile co-deposition platform, which enables the fabrication Ni–$C_{60}$ composite films under well-controlled conditions and their subsequent modification through multiple external stimuli, allowing a direct and meaningful comparison of stimulus-driven morphological and structural evolution. The resulting nanostructures produced on the surface of silicon crystals were subsequently modified through several distinct methods and thoroughly characterized. Well-established ion beams analytical techniques, i.e., Rutherford Backscattering Spectrometry (RBS), were used to determine the films' initial composition and thickness. It should be noted that the absence of pure Ni and pure $C_{60}$ control samples represent a known limitation of the present study. However, the structural evolution of Ni thin films under thermal or irradiation treatments is well known to be characterized by grain growth and surface oxidation, without phase separation or self-organized pattern formation [28, 29]. Similarly, pure fullerene films are known to undergo polymerization or amorphization under thermal, laser, or ion irradiation [30]. These behaviors differ qualitatively from the stimulus induced morphological evolution observed in the hybrid system, where phase coexistence and interfacial instability enable additional evolution pathways. Both the "as-deposited" and modified samples exhibited significant morphological and structural evolution in response to the applied external stimuli. The combination of Raman and SEM anal-yses support the conclusion that annealing procedure produces a phase separation between C and Ni and from $C_{60}$ amorphous carbon is formed. Instead, laser irradiation produces aligned nanoparticles and graphite crystallization of the C phase. From a morphological point of view, the ion irradiation does not change mostly the surface of the samples, but increases the amorphous phase of the samples, as shown in Raman results. An important result is exhibited in electrical characteristics, in which the produced films are not strongly affected by the external stimuli maintaining their sheet resistance and only slightly modified. The annealed sample represents the main exception, as thermally induced phase separation disrupts the conductive pathways, leading to a pronounced increase in effective resistance.

**Author Contributions:** Conceptualization, G.C. and J.V.; methodology, G.C., Y.K., S.V., E. S. and K.T.; validation, R.M., J.V.; formal analysis, G.C., J.N. and S.V.; investigation Y.K., J.N., S.V. and E.S.; resources, G. C. and K.T.; data curation, G.C., R.M., S.V.; writing—original draft preparation, G.C. and S.V.; writing—review & editing, G.C. and S.V.; project administration, G.C. and K.T.; funding acquisition, G.C. and K.T. All authors have read and agreed to the published version of the manuscript.

**Funding:** The authors acknowledged the support of Czech Academy of Science Mobility Plus Project, Grant No. JSPS-24-12 and JSPS Bilateral Program Number JPJSBP120242501. Measurements were carried out at the CANAM infrastructure of the NPI CAS Rez under project LM 2015056.

**Data Availability Statement:** The data presented in this study are available on request from the corresponding author. (The data supporting this study are not openly available as they are part of ongoing research.)

**Conflicts of Interest:** The authors declare no conflicts of interest.